\documentclass[%
 reprint,
 amsmath,amssymb,
 aps,
]{revtex4-2}

\usepackage{graphicx}
\usepackage{dcolumn}
\usepackage{bm}

\usepackage{caption}
\usepackage{subcaption}
\usepackage{comment}
\usepackage{xcolor}
\usepackage{siunitx}
\usepackage{braket}
\usepackage{xspace}

\newcommand{\HCI}[2]{#1$^{#2+}$\xspace}
\newcommand{\Be}[0]{\HCI{Be}{}}
\newcommand{\threePzero}[0]{$^3$P$_0$\xspace}
\newcommand{\threePone}[0]{$^3$P$_1$\xspace}
\newcommand{\threePtwo}[0]{$^3$P$_2$\xspace}

\begin{document}

\title{Identification of highly-forbidden optical transitions in highly charged ions}

\author{Shuying Chen$^{1,*}$, Lukas J. Spieß$^{1,*}$, Alexander Wilzewski$^1$, Malte Wehrheim$^1$, Kai Dietze$^1$, Ivan Vybornyi$^2$, Klemens Hammerer$^2$, José R. Crespo López-Urrutia$^3$, and Piet O. Schmidt$^{1,4}$}

\affiliation{$^1$QUEST Institute for Experimental Quantum Metrology, Physikalisch-Technische Bundesanstalt, Bundesallee 100, 38116 Braunschweig, Germany }
\affiliation{$^2$Institute of Theoretical Physics, Leibniz Universität Hannover,  Appelstrasse 2, 30167 Hannover, Germany}
\affiliation{$^3$Max-Planck-Institut für Kernphysik, Saupfercheckweg 1, 69117 Heidelberg, Germany }
\affiliation{$^4$Institut für Quantenoptik, Leibniz Universität Hannover, Welfengarten 1, 30167 Hannover, Germany}
\email{shuying.chen@quantummetrology.de}
\email{lukas.spiess@quantummetrology.de}
\vspace{10pt}

\begin{abstract}
Optical clocks represent the most precise experimental devices, finding application in fields spanning from frequency metrology to fundamental physics. 
Recently, the first highly charged ions (HCI) based optical clock was demonstrated using Ar$^{13+}$, opening up a plethora of novel systems with advantageous atomic properties for high accuracy clocks. While numerous candidate systems have been explored theoretically, the considerable uncertainty of the clock transition frequency for most species poses experimental challenges. 
Here, we close this gap by exploring quantum logic-inspired experimental search techniques for sub-Hertz clock transitions in HCI confined to a linear Paul trap. These techniques encompass Rabi excitation, an optical dipole force (ODF) approach, and linear continuous sweeping (LCS) and their applicability for different types of HCI. Through our investigation, we provide tools to pave the way for the development of exceptionally precise HCI-based optical clocks.
\end{abstract}

%
\vspace{2pc}
%
%
\maketitle
%
%

\section{Introduction}
Highly charged ions (HCI) represent a class of atoms that are characterized by the absence of multiple electrons from their electron shells \cite{kozlov_hci_2018}. The remaining electrons are bound more tightly to the atomic nuclei and exhibit extreme properties that distinguish HCI from neutral atoms or singly charged ions. Notably, several optical transitions within HCI have been identified that surpass neutral or singly charged atoms in their sensitivity to variations of the fine-structure constant \cite{kozlov_hci_2018, yu_highly_2023, berengut_enhanced_2010,berengut_testing_2013, berengut_electron-hole_2011, berengut_optical_2012, dzuba_actinide_2015, windberger_identification_2015, porsev_optical_2020}. Furthermore, the electronic transitions within HCI are inherently less sensitive to the influence of external electric and magnetic fields, resulting in smaller shifts of their transition frequencies when compared to conventional systems \cite{kozlov_hci_2018, schiller_optical_2007, derevianko_highly_2012, berengut_highly_2012}. This makes HCI good candidates for frequency references in optical atomic clocks \cite{kozlov_hci_2018, yu_highly_2023,yu_selected_2018, beloy_quadruply_2020, beloy_prospects_2021,  allehabi_atomic_2022, dzuba_cm15_2023}. We recently demonstrated that this is indeed experimentally feasible and not just an elusive theoretical possibility by demonstrating the first of such clocks based on \HCI{Ar}{13} \cite{king_optical_2022}.

However, this clock is limited by the relatively short lifetime of the excited state in the clock transition of about \SI{10}{\milli\second} which results in a comparatively large instability above $10^{-14}$ at \SI{1}{\second} \cite{peik_laser_2006}. To fully unlock the potential of HCI-based optical clocks, HCI with optical transitions offering excited-state lifetimes beyond \SI{1}{\second} are needed. Conversely, a long lifetime implies a low scattering rate of photons on that transition which makes their experimental detection challenging.

For transitions with lifetimes of up to \SI{100}{\milli\second}, fluorescence detection of a HCI plasma trapped in electron beam ion traps (EBITs) is an established technique. Numerous optical transitions have been detected with uncertainties below \SI{1}{\giga\hertz} \cite{mackel_laser_2013, rehbehn_sensitivity_2021, crespo_high_2005, draganic_high_2003}. From there, coherent laser excitation in a Paul trap using quantum logic spectroscopy (QLS) \cite{schmidt_spectroscopy_2005} is possible to achieve uncertainties at the \SI{}{\hertz}-level \cite{micke_coherent_2020} enabling their use for optical clocks. 

Optical transitions in HCI from electronic states with a lifetime exceeding \SI{1}{\second} to the ground state could only be inferred in few cases using the Ritz-Rydberg combination principle \cite{bekker_detection_2019, rehbehn_narrow_2023}. In the XUV range, ultra-precise mass measurements in Penning traps allowed to determine the mass differences between very long-lived metastable states (days) from their respective ground states, and thus identification of potential clock transitions \cite{schussler_detection_2020, kromer_observation_2023}, but for most cases though, such a detection has not been demonstrated. Alternatively, \textit{ab-initio} atomic structure calculations with an uncertainty on the order of \SI{}{\tera\hertz} can guide experimental searches using laser spectroscopy. Locating an optical transition of a sub-hertz natural linewidth within such a large frequency interval is a major obstacle in realising the next-generation HCI clocks. Comparable challenges are also present for other systems including neutral atoms, singly charged ions, molecules or nuclear transitions \cite{peik_nuclear_2021, safronova_search_2018}.

In this paper, we discuss several methods for the identification of highly forbidden transitions in HCI within experimentally available resources and feasible time scales. They are based on QLS techniques, 
in which a single HCI is co-trapped with a so-called logic ion. The logic ion provides sympathetic cooling, state preparation and readout of the HCI's clock state via shared motional modes \cite{schmidt_spectroscopy_2005}.
 
\begin{figure*}[t]
    \centering
    \includegraphics[width=0.75\textwidth]{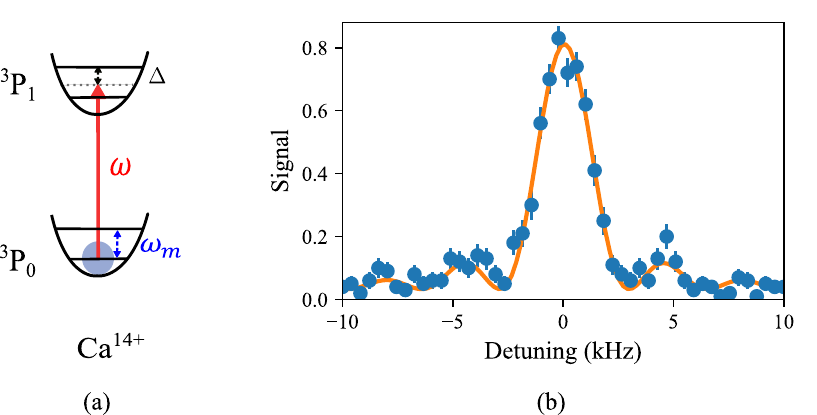}
    \caption{(a) Blue sideband Rabi excitation scheme for a \HCI{Ca}{14} - \Be two-ion crystal. A clock beam with frequency $\omega$ (in red) couples the ground state $|g\rangle =$ \threePzero\  to the excited clock state $|e\rangle =$ \threePone\ on the first-order blue sideband with detuning $\Delta$, where $\omega_m/(2\pi) =$ \SI{1.5}{\mega\hertz} represents the motional frequency. After simultaneous electronic and motional excitation, the shared motional excitation can then be detected on the co-trapped \Be, which is not shown here.  (b) A blue sideband excitation line profile of \HCI{Ca}{14}. Blue dots are experimental data and error bar are from quantum projection noise from 100 experimental repetitions. Orange solid line is from a sinc funciton fitting.}
    \label{fig:rabi}
\end{figure*}

This encompasses three methods. In the first two schemes, named "blue sideband Rabi excitation" and "optical dipole force (ODF)" \cite{hume_trapped-ion_2011,wolf_non-destructive_2016},  
the electronic excitation of the clock transition couples directly to the motional state. The motional excitation depends on the laser detuning, which is used to identify the weak transition within a broad range. The generated phonon is then detected on the logic ion.
The third method, "linear continuous sweeping (LCS)" of the laser frequency is used to excite the transition between electronic ground state and clock state via a rapid adiabatic passage (RAP)-like excitation process \cite{zener_non_1932}.
After exciting the electronic state, a known auxiliary transition from the ground state (denoted as the logic transition) is utilized to generate phonons in the shared motional mode, only if the HCI has not been excited to the upper clock state, akin to electron shelving \cite{nagourney_shelved_1986, dehmelt_stored_1982}. The generated phonon is then also detected on the logic ion. For the purpose of transition identification, perfect population transfer is not necessary. Instead, a sufficient signal is required to distinguish an effective excitation event from background noise.

The paper follows this structure: Section 2 explains the principles and demonstrates the experiments of the three methods. Section 3 analyses the efficiency of the search with different methods, as well as the limitations and challenges for application. Finally, Section 4 provides a summary and outlook.

\section{Searching Methods}

\subsection{Blue sideband Rabi excitation \label{sec:rabi}}

%
This method is a simplified version of QLS \cite{schmidt_spectroscopy_2005}.
For this, a two-ion system consisting of a HCI and a logic ion (\Be here) are co-trapped in a linear Paul trap. 
Figure~\ref{fig:rabi}(a) shows the involved energy levels of the HCI, \HCI{Ca}{14} used here as an example, as well as the required laser to drive the blue sideband Rabi excitation in the HCI. The logic ion \Be is not shown in the graph for simplicity, but the energy levels can be easily found in previous publications \cite{king_optical_2022, micke_coherent_2020, king_algorithmic_2021}. Initially the two ion crystal is prepared in the motional ground state of an axial mode.
A clock beam at variable frequency $\omega$ couples the ground state \threePzero\ to the clock state \threePone\ in \HCI{Ca}{14} on the BSB when $\omega = \omega_0 + \omega_m$, where $\omega_0$ is the purely electronic transition frequency (carrier) and $\omega_m$ is the motional frequency. The phonon is then detected and read out on the \Be in a quantum logic-like sequence (not shown in Figure~1(a)) \cite{ king_optical_2022, micke_coherent_2020,schmidt_quantum_2009}. Figure~1(b) shows a line profile of \HCI{Ca}{14} blue sideband Rabi excitation. Blue dots are experimental data while error bars are calculated quantum projection noise from 100 experimental repetitions. The excitation probability signal for this process with a rectangular-shaped laser pulse of length $t_p$ is given by the well-known Rabi formula:
  \begin{equation}
    P\textsubscript{Rabi} = \left(\frac{ \eta \Omega_\mathrm{} t_p }{2}\right)^2 \mathrm{sinc} ^2 \left( \frac{\Omega_\mathrm{R} t_\mathrm{p}}{2} \right),
    \label{eq:rabi_line}
\end{equation}
where $\Omega$ represents the on-resonance Rabi frequency of the carrier transition and $\Omega_\mathrm{R} = \sqrt{\eta^2\Omega^2+\Delta^2}$ is the effective Rabi frequency with $\Delta = \omega - (\omega_0 + \omega_m)$ being the detuning of the laser frequency from the BSB resonance frequency. $\eta$ is the Lamb Dicke parameter \cite{leibfried_quantum_2003}. The linewidth of Eq. (\ref{eq:rabi_line}) is approximately given by $\Omega$ for a probe time $t_{\mathrm{p}}\geq \pi/\Omega$. The orange solid line in Figure~\ref{fig:rabi}(b) is a fitting form Eq.~(\ref{eq:rabi_line}). 

\begin{figure*}[t]
    \centering
   \includegraphics[width=0.9\textwidth]{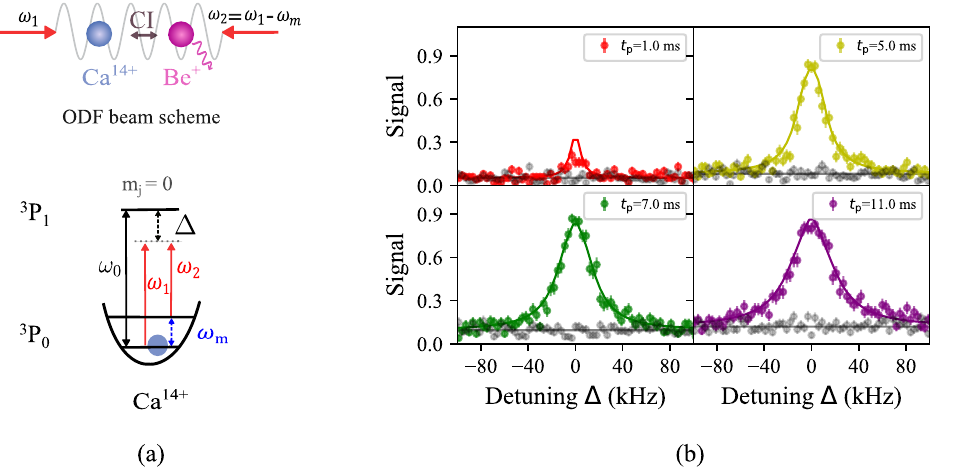}
    \caption{(a) The upper plot displays the ODF beam scheme, while the lower graph shows the involved energy level of \HCI{Ca}{14}. Two beams, with a frequency difference of $\omega_1-\omega_2=\omega_m$, are applied in counter-propagating directions to the ion. The lasers couple the ground state \threePzero\ with the clock state \threePone. A \SI{23}{\micro\tesla} magnetic bias field splits the Zeeman sublevels of the $^3\mathrm{P}_1$ by \SI{500}{\kilo\hertz}. Only the $m_j = 0$ sublevel in the excited state \threePone\ is shown here. CI : Coulomb interaction. (b) The ODF spectrum profiles versus detuning with different probe times are shown with different colours. Experimental data is represented by circles, while solid lines indicate theory fitting. Black circles depict the background signals from motional heating when the ODF beams are switched off. The black solid line is the predicted value of the background from Eq.~(\ref{pt}) in the main text. }
    \label{fig1}
\end{figure*}

To increase the transition search efficiency for Rabi excitation, a sequence of $N_{\mathrm{p}}$ probe pulses with varying frequency of step $\delta f$ can be implemented after initialising the motional mode to the ground state. This reduces the overall experimental preparation and readout time per frequency step. The sequence is repeated $N_{\mathrm{c}}$ times with identical parameters to decrease quantum projection noise \cite{itano_quantum_1993}. 

This method was previously used for searching of the magnetic-dipole allowed transitions in \HCI{Ar}{13} and \HCI{Ca}{14}. For these, a closed transition ($m_J \rightarrow m_{J\pm1}'$) was present. Driving these transitions allows the HCI to return to the same Zeeman substate with high probability if spontaneous decay occurs. 
Additionally, their excited-state lifetime of \SI{10}{\milli\second} is shorter than the time required to prepare the two-ion crystal such that no deexcitation of the excited state needed to be performed. Neither of these properties is generally present in the transitions being sought in HCI, which complicates transition searches using Rabi excitation.

If the excited state has a short lifetime, e.g. \SI{10}{\milli\second}, but is not closed, decay to a different ground state rather than the initial one is likely, again requiring some form of ground-state preparation. Further, if the excited-state lifetime is over \SI{1}{\second}, optical pumping from the excited-state to the ground state is essential. Otherwise, the electron will stay in the excited state for extended periods of time, resulting in the failure of BSB excitation. Optical pumping can be achieved through quantum-logic assisted optical pumping \cite{ schmidt_spectroscopy_2005,micke_coherent_2020}, though this is challenging due to imperfect initial knowledge of the Rabi frequency and shifts induced by the probe laser. This is not a fundamental limitation but likely reduces fidelity.

%

\subsection{Optical dipole force}
This section introduces a method that overcomes some of the limitations of Rabi excitation by employing a detuned dispersive force, instead of direct electronic excitation. Through this it is less sensitive to the inner level structure of the involved states and eliminates the necessity of electronic state preparation, since ideally the atom always remains in the same state.

Figure~\ref{fig1}(a) shows the lasers required for its implementation and the HCI level structure, here \HCI{Ca}{14} is used as an example. Two counter-propagating lasers with frequencies $\omega_1$ and $\omega_2$ create a moving optical lattice on the trapped ions. Setting their frequency difference equal to one of the motional mode frequencies $\omega_1-\omega_2=\omega_m$, exerts an optical dipole force (ODF) on the ion crystal \cite{wolf_non-destructive_2016, hume_trapped-ion_2011}. In the far-off resonant scenario, the ODF implements a displacement operator on the motional state. It transforms a ground state-cooled ion crystal from motional state $\ket{n=0}$ to a coherent state $\ket{\alpha}$ with Fock number probability distribution 
\begin{equation}
    P_\alpha(n) = e^{-\vert\alpha\vert^2} \frac{\vert\alpha\vert^{2n}}{n!},
\end{equation}
where $|\alpha| = \frac{\eta t_p \Omega_1 \Omega_2}{4\Delta}$ is the displacement amplitude, $\Omega_1$ and $\Omega_2$ represent the on-resonant Rabi frequencies of the driving light fields at $\omega_1$ and $\omega_2$ respectively, and $t_p$ is the pulse length. The Fock state and Lamb Dicke parameter are denoted by $n$ and $\eta$, respectively. The detuning of the laser $\omega_1$ to the resonance of the atomic transition $\omega_0$ is $\Delta = \omega_1-\omega_0 = \omega_2 + \omega_m -\omega_0$. The motional excitation can then be read out on the \Be ion. With rapid adiabatic passage \cite{king_optical_2022, gebert_corrigendum:_2018, gebert_detection_2016} on the RSB of the \Be transition, all population outside the motional ground state contributes to the signal, thus the readout signal is written as
\begin{equation}
    P\textsubscript{ODF} = 1-P_\alpha(0) = 1- \exp\left[-\left(\frac{\eta t_p \Omega_1 \Omega_2}{4\vert\Delta\vert}\right)^2\right].
    \label{eq:p_odf}
\end{equation}
Setting $P\textsubscript{ODF}(\Delta_{1/2})=0.5$, we find the signal's width to be
\begin{equation}
    \Delta_{1/2}=\frac{\eta t_p \Omega_1 \Omega_2}{4 \sqrt{\ln(2)}},
\label{delta_odf}
\end{equation}
with $\Delta_{1/2}$ being proportional to the probe time $t_p$.\

For an experimental demonstration of this method, we again used the \HCI{Ca}{14}-\Be\ system. The level structure of \HCI{Ca}{14} comprises the \threePzero\ ground state and the \threePone\ excited state. A \SI{23}{\micro\tesla} magnetic field introduces a \SI{500}{\kilo\hertz} Zeeman splitting between the sub-levels ($m_j = -1, 0, 1$) of the \threePone\ excited state. For simplicity, only the $m_j=0$ level is shown in Figure~\ref{fig1}(a). The \threePzero ground and \threePone excited clock states are coupled by the two lattice light fields.
The first ODF beam ($\omega_1$) is aligned with the symmetry axis of the two-ion crystal and thus has unit projection to the axial mode. Due to geometric constraints of our experimental apparatus, the second beam ($\omega_2$) is delivered at a $15^\circ$ angle relative to the symmetry axis. This results in a slightly weakened coupling strength with the motional sideband, reduced to a factor of $\cos{15^\circ}\approx 0.97$. We verify the modelled signal of Eq.~(\ref{eq:p_odf}) by setting the on-resonance carrier Rabi frequencies $\Omega_{1,2}/2\pi \approx \SI{5}{\kilo\hertz}$, so that they are significantly smaller than the Zeeman splitting of approximately \SI{500}{\kilo\hertz}.
This allows to fulfill the far-off-resonance condition $\Omega_{1,2} \ll \Delta$ and to model \HCI{Ca}{14} as a two-level system. 

Figure~\ref{fig1}(b) shows the readout signal profile for a range of probe times $t_p$ with varying detuning $\Delta$. The coloured circles represent experimental excitation data, while the black circles indicate the background from motional heating when the ODF beams are off. 
The error bars on the data points are calculated from the quantum projection noise of the 100 experimental repetitions.

Firstly, the heating background in Figure~\ref{fig1}(b) visibly rises as the probe time increases. This is attributed to anomalous ion heating that drives the Fock state population $P(n)$ to a thermal distribution \cite{leibfried_quantum_2003}:
\begin{equation}
    P_{\bar{n}(t)}(n)=\frac{\bar n(t) ^n}{(\bar n(t) +1)^{n+1}}.
    \label{pn}
\end{equation}
The phonon excitation from the motional ground state caused by heating is described by the equation $\bar{n}(t)=n_0+\beta t$, resulting in a background signal following 
\begin{equation}
    P\textsubscript{thermal}(t)=1 - P_{\bar{n}(t)}(0)=1-\frac{1}{1+\beta t_p + n_0}.
    \label{pt}
\end{equation}
Here, $\beta$ denotes the heating rate, and $n_0$ represents the initial phonon number after sideband cooling. For our system, we measured $\beta=8.2$ 1/s and $n_0 = 0.05$ with sideband thermometry \cite{monroe_resolved-sideband_1995}. With this, Eq.~(\ref{pt}) describes the experimental background well, which is shown as the black solid line in the Figure~\ref{fig1}(b). 

\begin{figure}[h!]
\centering
\includegraphics[scale=0.8]{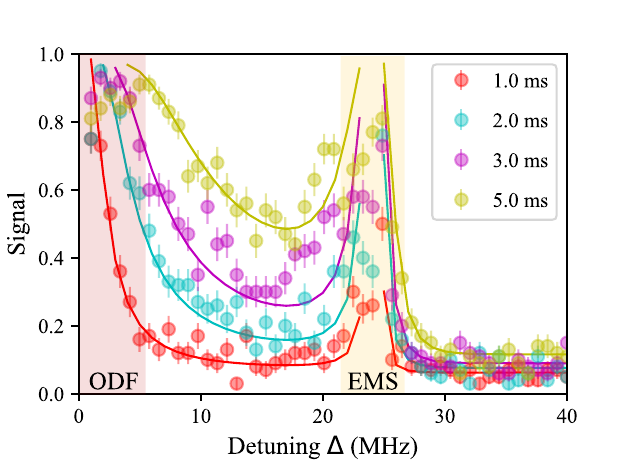}
\caption{ODF excitation in Ca$^{14+}$ with a larger resonant Rabi frequency of \SI{100}{\kilo\hertz}. The $y$-axis represents excitation probability and the $x$-axis frequency detuning. Experimental data with different probe times is indicated by round dots. Solid lines are simulations from solving the master equation (for details of the simulation, see Supplemental Material \cite{supplement}). Near vanishing detuning, denoted by a pink band, there is an near-resonant excitation peak. Additionally, at a detuning of around \SI{24}{\mega\hertz}, marked with a yellow band, there is another excitation peak from excess micromotion (EMS) from the trap rf drive field. The resonances are excluded from the simulation. Uncertainties result from quantum projection noise.}
\label{fig2}
\end{figure}

\begin{figure*}[t]
    \centering
    \includegraphics[width=0.75\textwidth]{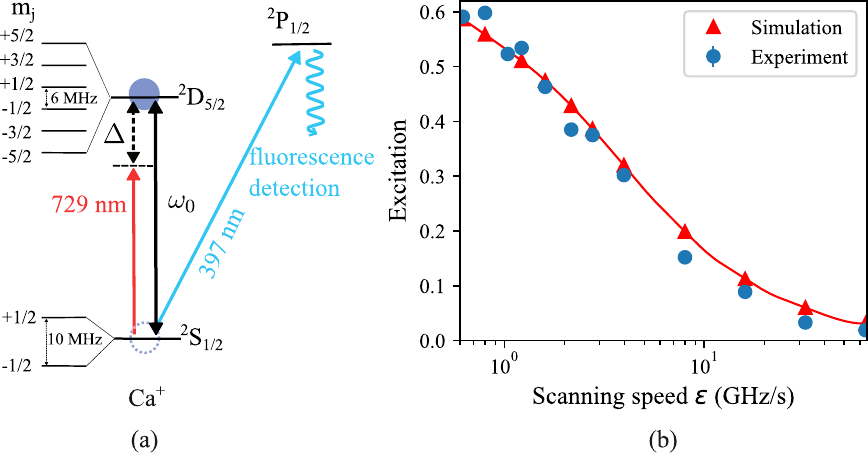}
    \caption{(a) Energy levels of Ca$^+$ including Zeeman sublevels and the lasers involved for LCS excitation and state detection. (b) LCS excitation spectrum versus scanning speed $\varepsilon$. The blue dots are experimental data while the red triangles are from numerical simulations based on a master equation (for details of the simulation, see Supplemental Material \cite{supplement}), with the connected red line as a guide to the eye. }
    \label{fig3}
\end{figure*}
The maximum amplitude of the ODF excitation probability, as illustrated in Figure~\ref{fig1}(b), increases with probe time up to \SI{7}{\milli\second}, where it saturates at approximately 0.87. This non-unity amplitude is caused by imperfect cooling, state preparation, and readout of the \Be ion.  Furthermore, the linewidth of the ODF excitation signal increases with the probe time, although we observe a decreased broadening when the probe time exceeds \SI{10}{\milli\second}, which we attribute to dephasing due to coupling of the motional state to the environment \cite{leibfried_quantum_2003, turchette_decoherence_2000,fluhmann_encoding_2019, intravaia_quantum_2003, simon_measuring_2020}. Experimentally, we also investigated the relative phase noise between the two laser beams and found no dependence of the signal on the phase noise level.
The colored solid lines are simulations solving the Lindblad master equation accounting for the optical dipole force and relevant dephasing and heating process (for details of the simulations, see Supplemental Material \cite{supplement}) \cite{leibfried_quantum_2003, turchette_decoherence_2000, fluhmann_encoding_2019, intravaia_quantum_2003, simon_measuring_2020, meekhof_generation_1996}. These effects cause degradation of the ODF signal for longer probe times and describe the experimental data well, indicating that these effects limit the probe time. They also impact the signal-to-noise ratio of ODF excitation for different probe times differently, which in turn affects the efficiency of line searching. Section \ref{sec3} provides an analysis of the optimal probe time that can be used to minimize the line search time.

With a higher resonant Rabi frequency of $\Omega/2\pi\approx \SI{100}{\kilo\hertz}$, all three Zeeman components ($m_j=-1, 0, 1$), with Zeeman splitting of around \SI{500}{\kilo\hertz} 
in the excited state \threePone 
collectively contribute to the overall ODF excitation probability. For $\Omega/2\pi\approx \SI{100}{\kilo\hertz}$ an excitation signal with two discernible resonant peaks is observed in Figure~\ref{fig2} for different probe times. The peak near \SI{0}{\mega\hertz} detuning corresponds to the resonant situation when $\omega_1=\omega_0$ in the HCI. The peak near \SI{24}{\mega\hertz} arises from excess micromotion present in the employed trap \cite{leopold_cryogenic_2018}. The contributions from the two transitions interfere with an intrinsic relative phase of $\pi$ between the carrier and micromotion sideband, in addition to a phase that depends on the detuning. In the interval between the resonances, the relative detuning phase with respect to the two resonances is $\pi$, and the two contributions interfere constructively, which enhances the signal. Beyond the micromotion sideband, there is no relative detuning phase; only the relative phase of the carrier and the micromotion sideband of $\pi$ is present, which results in destructive interference and a reduction in excitation. The solid lines are simulations from solving master equations numerically as mentioned above, but including the three Zeeman sublevels in \threePone and the micromotion excitations. The excitation data with \SI{3}{\milli\second} probe time (pink dots) is slightly deviating from the simulation curve near \SI{24}{\mega\hertz} detuning. We attribute this to an additional heating from fluctuating patch potentials or rf noise during that measurement.

Again, the effective linewidths of the ODF excitation near the resonances increase with longer probe times, but with additional enhancement from the interference effect. For a probe time of \SI{3}{\milli\second}, the excitation spectrum has a linewidth near \SI{10}{\mega\hertz}, which equals 100 times the resonant Rabi frequency. 

Notably, in the far-detuned scenario, the ODF preserves the initial state, making it suitable for situations where electronic excitation needs to be avoided. This is particularly relevant for short excited-state lifetimes, where spontaneous decay can lead to population transfer between multiple ground states \cite{wolf_non-destructive_2016}. 

\subsection{Linear Continuous Sweep}

In a two-level system, one can coherently populate the excited state with almost unit fidelity by continuously scanning the laser frequency over the resonance of the atomic transition. This is possible when the frequency sweep rate $\varepsilon \ll  \Omega^2$ with the on-resonance Rabi frequency $\Omega$. For optimal population transfer, we furthermore require the laser linewidth $\gamma  \ll  \Omega$  \cite{lacour_uniform_2007, noel_adiabatic_2012},  so that the ground-state population is adiabatically transferred to the excited state. This rapid adiabatic passage (RAP) scheme was used before to search for atomic transitions \cite{furst_coherent_2020} or coherent population transfer \cite{wunderlich_robust_2007}. In those cases, the two-level approximation is valid since the frequency sweep range before read-out and re-initialization of the system was much smaller than the Zeeman splitting of the involved manifolds. For the cases considered here, this is not generally true because one sweep cycle could scan over the resonances of all Zeeman transitions, making the dynamics of the atomic population more complex. However, motivated by the RAP's robustness and simple experimental implementation for a two-level system, we here investigate this method for a multi-level system and a large sweep range, which covers all ground and excited states, named linear continuous sweep (LCS). 

We are mainly interested in two things. Firstly, how robust the detection of atomic excitation after a laser frequency sweep is, when the on-resonance Rabi frequency $\Omega$ and laser linewidth $\gamma$ are poorly known or, in other words, how sensitive the scheme is to the scanning speed $\varepsilon$ for a given $\Omega$ and $\gamma$. Secondly, what overall scanning speeds one can achieve with realistic values for $\Omega$ and $\gamma$.

We tested this experimentally on the  $^2\mathrm{S}_{1/2} \rightarrow ^2\mathrm{D}_{5/2}$ transition at \SI{729}{\nano\meter} in \HCI{Ca}{} (see Figure~\ref{fig3}(a)) in the following way. After Doppler cooling on the \SI{397}{\nano\meter} transition the ion is optically pumped to the $^2\mathrm{S}_{1/2}, m_J=-1/2$ state. Then, the \SI{729}{\nano\meter} laser frequency is scanned over the atomic resonances of all Zeeman transitions having a center frequency $\omega_0$. We adjusted the laser's power and polarization such that the on-resonance Rabi frequencies $\Omega$ of all the Zeeman transitions ranged from $\SI{1}{\kilo\hertz}$ to $\SI{10}{\kilo\hertz}$ to be comparable with those of the methods discussed in the previous sections and match realistic values in an actual line search scenario. The \SI{729}{\nano\meter} laser source is an external cavity diode laser (ECDL). Applying a linear voltage ramp over a period $T$ to the piezoelectric actuator that controls the grating angle of the ECDL, results in varying detuning $\Delta(t) = \omega_0 +  \varepsilon t$ with $t \in [-T/2,+T/2]$. During the frequency ramp, the laser is free-running without frequency-stabilization to any external reference, such as an optical cavity, and has a linewidth of $\gamma/2\pi \approx \SI{100}{\kilo\hertz}$ according to the manufacturer's specifications. The total frequency range covered in one scan $\Delta(+T/2) - \Delta(T/2)$ is much larger than the splitting of the individual Zeeman components: A magnetic field of \SI{100}{\micro\tesla} splits the ground state levels by \SI{10}{\mega\hertz} and the excited state levels by \SI{6}{\mega\hertz}, while the total frequency scan range is at least \SI{60}{\mega\hertz}. After laser sweeping, the excited state population is read out by detecting fluorescence (or absence thereof) from the \SI{397}{\nano\meter} transition, as shown in Figure~\ref{fig3}(b).

The blue circles in Figure~\ref{fig3}(b) represent the population transferred to the excited states for varying scanning speeds $\varepsilon$. Up to the scanning speed of \SI{1}{\giga\hertz\per\second}, the excitations are more than 50\% of the population with a maximum of about 60\%. It decreases for $\varepsilon > \SI{1}{\giga\hertz\per\second}$ and falls below 10\% for $\varepsilon > \SI{10}{\giga\hertz\per\second}$. For the theoretical description of these measurements, a modified Landau-Zener equation \cite{lacour_uniform_2007} is not applicable for two reasons. First, it assumes a two-level system like in \cite{furst_coherent_2020}. Secondly, we explore a parameter range where the adiabaticity criteria, $\varepsilon \ll \Omega \cdot \gamma$ and $\gamma \ll \Omega$, are not strictly fulfilled.
\begin{figure*}[t]
    \centering
    \includegraphics[width=0.75\textwidth]{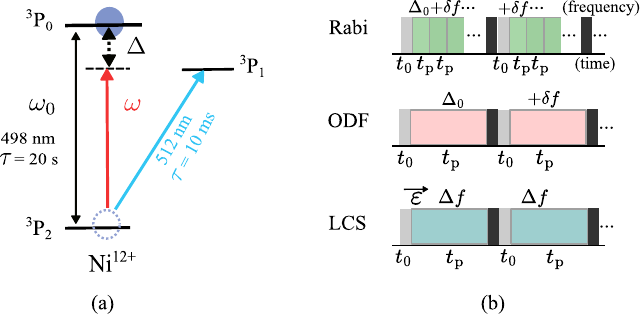}
    \caption{(a) \HCI{Ni}{12} level structure with clock transition \threePtwo $\rightarrow$ \threePzero and logic transition \threePtwo $\rightarrow$ \threePone. 
    (b) The time sequences for the \HCI{Ni}{12} search with the three methods introduced above. The x-axis is time. The grey rectangle blocks are state preparation and black blocks are state readout. These two overhead processes within a cycle take a total time of $t_0$. Green blocks are probe pulses for the Rabi method with $N_{\mathrm{p}}$ pulses within each cycle, and with a frequency difference of $\delta f$ between the pulses.
    The pink blocks are probe pulses for ODF and the frequency between two pulses is marked also as $\delta f$.
    The blue blocks are LCS with speed $\varepsilon=\Delta f/t_{\mathrm{p}}$, where $\Delta f$ and $t_{\mathrm{p}}$ are the scan range and sweep time, respectively. }
    \label{fig:Ni12+ level}
\end{figure*}
For these reasons, we numerically solve the master equation. In the simulations, an initial density matrix $\rho_0$ with all the population in the $^2\mathrm{S}_{1/2}, m_J=-1/2$ state is propagated with a time-dependent Hamiltonian containing the changing laser frequency and a set of Lindblad operators accounting for the finite laser linewidth $\gamma$ and spontaneous atomic decay $\Gamma$ (for details of the numerical simulation, see Supplemental Material \cite{supplement}). The red triangles in Figure~\ref{fig3}(b) are the results of numerical simulations taking the values of the experimental parameters for $\Omega$, $\Gamma$, $\gamma$ and $\varepsilon$ as inputs. The red line in Figure~\ref{fig3}(b) is a connection of the simulations and serves as a guide to the eye. The simulation results match the observed excited state population.

So far we have omitted the required level structure and the difference between a general HCI and \HCI{Ca}{}. As Figure~\ref{fig3}(a) shows, a detection of the excited-state population is needed as the dominant excitation of LCS is carrier excitation. Unlike \HCI{Ca}{}, HCI in general does not exhibit a transition for direct fluorescence detection.  Instead, quantum logic spectroscopy of another known logic transition sharing the same ground state is used. The excitation probability of the logic transition is mapped onto the logic ion where fluorescence detection is feasible. In this way, the excitation of the clock transition can be determined from the logic transition excitation readout. This means that, for LCS used in HCI, an additional known logic transition from the same ground state to transfer excitation of the HCI clock transition to \Be is required. When such a logic transition is available, the line excitation and readout will be similar as the LCS used in \HCI{Ca}{} here. 

\section{Search efficiency analysis} 
\label{sec3}

When searching for a narrow transition in a HCI, we need to detect a signal with a certain confidence level (here, we assume a 95$\%$ confidence level) in the shortest time possible. To determine the time required to scan through the initial uncertainty range, the effective search speed $s$, including all experimental overhead, is of interest. 

In the following, we will estimate $s$ for the three different methods discussed above and compare them. 
%
For the estimate of $s$, it is useful to think in terms of the \textit{experimental cycles}. For each method, an experimental cycle can be divided into three segments: (a) initial state preparation, (b) an attempt to excite the HCI with the lasers and (c) readout of the excited state population. 
Each experimental cycle covers a certain frequency range $\Delta f$ and has to be repeated $N_c$ number of times to separate the signal from the noise (here mainly QPN) and the background (here mainly anomalous heating of the ion). Denoting the time it takes to cover $\Delta f$ within $\Delta T$, the effective search speed is $s=\Delta f/\Delta T$. Also contained in $\Delta T$ is the overhead time $t_0$ required for state preparation and read-out.

For the estimates and comparison below, we conservatively assume an overhead time of $t_0 = \SI{10}{\milli\second}$ for the Rabi and ODF methods, and $t_0 = \SI{20}{\milli\second}$ for the LCS method, because signal readout via the logic transition requires additional time. The protocol for the LCS readout includes an additional quantum non-demolition (QND) measurement step [55] on the logic transition to detect the ground state population and generate motional excitation. This QND measurement can be achieved by applying a far-off resonant ODF to the logic transition without destroying the ground state, and can thus can be repeated many times to increase the readout fidelity to near-unity. The generated phonon can then be detected on Be+ using QLS [18,25,38]. Furthermore, we consider a transition with an excited-state lifetime of several seconds, e.g. the \threePtwo\ $\rightarrow$ \threePzero\ transition in \HCI{Ni}{12}, shown in Figure~\ref{fig:Ni12+ level}(a). For such a transition, an on-resonance carrier Rabi frequency of $\Omega/2\pi = \SI{5}{\kilo\hertz}$ is realistic and will be used in the following. As shown in Figure~\ref{fig:Ni12+ level}, the clock transition is an electric quadrupole (E2) transition from \threePtwo\ $\rightarrow$ \threePzero\ with a predicted lifetime of \SI{20}{\second}. The \threePtwo\ $\rightarrow$ \threePone\ logic transition has a predicted lifetime of around \SI{6.5}{\milli\second} \cite{yu_selected_2018, liang_probing_2021}. We assume a Lamb-Dicke parameter of $\eta=0.1$, similar to what we employed for \HCI{Ar}{13} and \HCI{Ca}{14}.

For the Rabi method, the probe time $t\textsubscript{p}$ is fixed to be identical to the time required for a $\pi$-rotation of the BSB transition, i.e. $t\textsubscript{p} = \pi/\eta\Omega$. We also assume ideal Rabi excitation, as described by Eq.~(\ref{eq:rabi_line}). Since $t\textsubscript{p}\ll t\textsubscript{0}$, we apply $N\textsubscript{p}$ pulses detuned by a frequency step $\delta f$ within an experimental cycle, instead of applying one single probe pulse. This process is then repeated $N\textsubscript{c}$ times to reduce quantum projection noise. The achievable effective search speed is
\begin{equation}
    s\textsubscript{Rabi} = \frac{\Delta f}{\Delta T} = \frac{\delta f\times N_\textsubscript{p}}{(t_{p} \times N\textsubscript{p}+t_0) \times N\textsubscript{c}}.
    \label{eq:rabi_eff}
\end{equation}
 \begin{figure*}[th]
    \centering
    \includegraphics[width=1\linewidth]{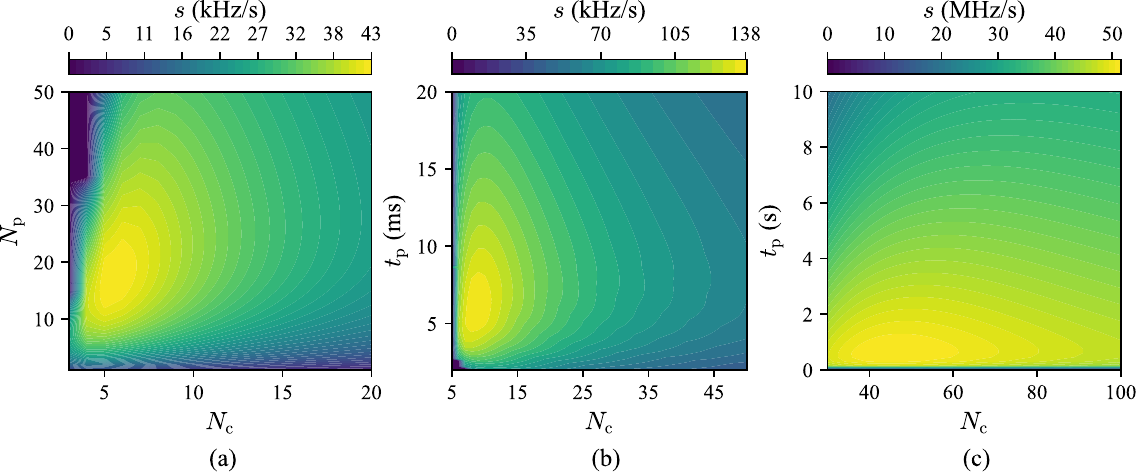}
    \caption{Effective search speed estimate for the three different methods. \HCI{Ni}{12} with an excited state lifetime of \SI{20}{\second} is used as a prototypical example. (a) Search time with Rabi BSB excitation, where $N_{\mathrm{p}}$ is the number of Rabi pulses within a cycle. (b) Search time with ODF excitation, where $t_{\mathrm{p}}$ is the probe time of each ODF pulse. (c) Search time with the LCS scheme, where $t_{\mathrm{p}}$ is the sweep time. Note the different scaling of the search speed. For all plots, the $x$ axis is the experimental cycle repeated times $N_{\mathrm{c}}$ and $s$ is the effective search speed. }
    \label{fig4}
\end{figure*}

The optimisation is done as follows: $N_\text{p}$ and $N_\text{c}$ are fixed initially. From this we then calculate the background excitation $P_\mathrm{bg}=P_\mathrm{thermal}(N_\text{p} t_\text{p})$ for the measurement time $N_\text{p} t_\text{p}$. From $N_\text{c}$ the QPN $\Delta P$ is calculated for the excitation $P$. We numerically optimise $\delta f$ such that, if scanning across the resonance, at least one measurement point gives an excitation probability $P_e$ which is with $95\%$ confidence above the background, i.e. 
\begin{equation}
    P_e - 1.95 \Delta P_e >  P_\text{bg} + 1.95 \Delta P_\text{bg}. 
    \label{95confidence}
\end{equation}
From a numerical search as shown in Figure~\ref{fig4}(a), we find a maximum effective speed for $s\textsubscript{Rabi}$ of \SI{43.1} {\kilo\hertz\per\second} with $N_\text{p} = 16$ and $N_\text{c} = 5$.

With similar considerations, we also find the maximum $s\textsubscript{ODF}$ for the ODF method. Here, only one probe pulse per experimental cycle is performed. The excitation profiles for ODF with varying probe time are produced from numerical simulations including the noise processes observed experimentally. The length of the probe pulse $t_{\mathrm{p}}$ determines the background excitation $P_\mathrm{bg}=P_\mathrm{thermal}(t_\text{p})$ and $N_{\mathrm{c}}$ determines the QPN. $\delta f$ is then optimised with similar process according to Eq.~(\ref{95confidence}). Therefore, the effective search speed evaluates to be
\begin{equation}
    s\textsubscript{ODF} = \frac{\Delta f}{\Delta T} = \frac{\delta f}{(t\textsubscript{p}+t_0) \times N\textsubscript{c}}.
\end{equation}
We optimise the effective speed $s\textsubscript{ODF}$ for varying $N_{\mathrm{c}}$ and $t_{\mathrm{p}}$ as shown in Figure~\ref{fig4}(b). The maximum effective search speed for ODF is found to be \SI{138.7}{\kilo\hertz\per\second} with $N_\text{c} = 9$ and $t_\text{p} =  \SI{5.8}{\milli\second}$. This is about three times higher than for Rabi excitation. Here, the main limitation is again the anomalous motional heating, which limits $t_{\mathrm{p}}$ and thus the effective speed. This is also not easy to overcome experimentally, as the heating rate of the Paul trap in use is already among the lowest ever reported \cite{king_algorithmic_2021}. 

The LCS method has an excitation probability numerically calculated depending on sweep speed $\varepsilon$ and laser linewidth $\gamma$ (for details of the calculation, see Supplemental Material \cite{supplement}). In practice, a Ti:saphire laser is more suitable for covering a larger frequency range without need to readjust. We assume $\gamma/2\pi= \SI{100}{\kilo\hertz}$, which is a typical linewidth achievable for a commercial Ti: saphire laser. The signal-to-noise ratio is limited by QPN of the excitation probability (see Figure~\ref{fig3}(c)) assuming the state detection itself can be performed with near-unity fidelity, for example, with QLS using a quantum non-demolition measurement \cite{hume_high-fidelity_2007}. That is why the overhead time of this method is set to \SI{20}{\milli\second}. The effective speed is
\begin{equation}
    s\textsubscript{LCS} = \frac{\Delta f}{\Delta T}= \frac{\Delta f}{(\frac{\Delta f}{\varepsilon} +t_0)\times N\textsubscript{c}},
\end{equation}
where $\Delta f$ is the frequency sweep range in a single scan, and is limited by the maximum mode-hop free tuning of the laser (here assumed to be \SI{20}{\giga\hertz} for a Ti:saphire laser). 
Additionally, the sweep time $t_{\mathrm{p}}$ is limited by spontaneous decay of the HCI. We optimise $\varepsilon$ and $\Delta f = t_{\mathrm{p}} \times \varepsilon$ for different $N_{\mathrm{c}}$ and scan time $t_{\mathrm{p}}$. The effective search speed as a function of $N_{\mathrm{c}}$, $\Delta f$ and $\varepsilon$ is shown in Figure~\ref{fig4}(c) with a maximum effective speed of about \SI{51.5}{\mega\hertz\per\second} taking the atomic decay into account at an optimum parameter set of $N_\text{c} = 46$, $t_\text{p} =  \SI{0.6}{\second}$ and $\varepsilon= \SI{2.4}{\giga\hertz / \second}$. 

A summary of the three methods is given in Table \ref{tab:efficiency}.
\begin{table}[h]
    \centering
    \begin{tabular}{ p{2cm}|p{2cm}|p{2cm}|p{2cm} } 
    \hline
                & Rabi excitation & ODF  &  LCS    \\ \hline    
    Optimal $s$ & \SI{43.1}{\kilo\hertz / \second}  &   \SI{138.7}{\kilo\hertz / \second}  & \SI{51.5}{\mega\hertz / \second}  \\ 
    Parameters & $t_\text{0} = \SI{10}{\milli\second}$ & $t_\text{0} = \SI{10}{\milli\second}$ & $t_\text{0} = \SI{20}{\milli\second}$ \\
    &$N\textsubscript{c}= 5$ & $N\textsubscript{c} = 9$ & $N\textsubscript{c} = 46$\\
    & $N\textsubscript{p}$ = 16  & $t\textsubscript{p} = \SI{5.8}{\milli\second}$ & $t\textsubscript{p}=$ \SI{0.6}{\second}\\
    & $\delta f = \SI{350}{\hertz}$ & $\delta f = \SI{19.8}{\kilo\hertz}$ & $\varepsilon$ = \SI{2.4}{\giga\hertz / \second} \\
    &$t\textsubscript{p} =$ \SI{6.2}{\milli\second} &   &$\Delta f$ = \SI{1.5}{\giga\hertz} \\
    Limitation & Heating & Heating & QPN \\\hline
    \end{tabular}
    \caption{Methods and corresponding optimal scanning parameters for HCI line searching (here e.g. \HCI{Ni}{12} clock transition). For all methods, a Rabi frequency of $\Omega/2\pi= \SI{5}{\kilo\hertz}$ was assumed.}
    \label{tab:efficiency}
\end{table}
In a real application, LCS can be used to enable robust and fast searches, provided a suitable logic transition is available \cite{bekker_detection_2019, yu_selected_2018, porsev_optical_2020}. If no logic transition is available, ODF is the best choice. 

\section{Conclusion}

We discussed techniques to experimentally search for ultra-narrow transitions in highly charged ions while dealing with significant frequency uncertainties using QLS-like techniques. 
For systems lacking a logic transition, ODF facilitates direct line searches via phonon excitation, which can then be read out by a logic ion. The estimated search speed is three times faster than Rabi excitation and more robust to spontaneous decay. The limitation is given by motional heating, competing with the signal. This technique is useful for identifying transitions where in-EBIT spectroscopy with \SI{}{\giga\hertz}-uncertainty is attainable and only a few hours of scanning are necessary.  
When a logic transition is available, linear continuous sweep (LCS) can be used to robustly search for ultra-narrow lines. This approach utilizes the known logic transition for electron shelving and quantum logic readout. Despite the \SI{5}{\tera\hertz}-uncertainty from ab-initio calculations, LCS would only require \SI{27}{\hour} to find a line with millihertz linewidth, corresponding to an excited state lieftime of seconds. In practice, additional time needs to be accounted to readjust the laser when scanning over such a large range after every mode-hop-free range. Diode lasers would frequently require changes to their operating parameters. Instead, a Ti:sapphire laser is more suitable for covering a larger frequency range, due to its large mode-hop-free range and the potential of automatic readjustment of its components. The achievable laser linewidth of commercial Ti:sapphire lasers is similar to that of a diode laser. 

With the combination of the aforementioned techniques, it is feasible to explore further clock transitions in HCI, including those with lifetimes of \SI{1}{\second}-\SI{100}{\second}. For instance, the 5f$_{5/2} \rightarrow$ 6p$_{1/2}$ clock transition in Cf$^{17+}$ has a lifetime of \SI{6}{\second} and has one of the highest known sensitivities to variations of the fine-structure constant \cite{dzuba_actinide_2015,porsev_optical_2020}. Together with the 5f$_{5/2} \rightarrow$ 5f$_{7/2}$ logic transition (excited-state lifetime of \SI{9}{\milli\second}) the LCS line search method proposed here would be suitable to cover the theoretical uncertainty of \SI{18}{\tera\hertz} \cite{porsev_optical_2020}. While the logic transition has a theoretical uncertainty of \SI{15}{\tera\hertz}, its coupling strength permits measurement in the EBIT to a precision of \SI{1}{\giga\hertz}, and detection with ODF is then feasible. 
A comparable approach could be employed in other atomic clock candidates such as \HCI{Pr}{9}, \HCI{Ba}{4}, and \HCI{Ge}{16} \cite{beloy_quadruply_2020,bekker_detection_2019,  allehabi_high-accuracy_2022}. The presented techniques are not restricted to HCI, but are generaly applicable to all trapped ion species that can be investigated using QLS.

\begin{acknowledgments}
All co-authors declare no conflicts of interest. 
The project was supported by the Physikalisch-Technische Bundesanstalt, the Max-Planck Society, the Max-Planck–Riken–PTB–Center for Time, Constants and Fundamental Symmetries, and the Deutsche Forschungsgemeinschaft (DFG, German Research Foundation) through SCHM2678/5-2, the collaborative research centres SFB 1225 ISOQUANT and SFB 1227 DQ-mat, and under Germany’s Excellence Strategy – EXC-2123 QuantumFrontiers – 390837967. The project 20FUN01 TSCAC has received funding from the EMPIR programme co-financed by the Participating States and from the European Union’s Horizon 2020 research and innovation programme. This project has received funding from the European Research Council (ERC) under the European Union’s Horizon 2020 research and innovation programme (grant agreement No 101019987). 
Data is available on \cite{chen_2024_11504689}.
\end{acknowledgments}
\textbf{Remark:} Following the submission, the aforementioned methods have been validated and successfully deployed to identify both the logic and clock transitions in \HCI{Ni}{12}. Further details will be included in an upcoming publication.
%

\clearpage

\appendix


\title{Supplementary: Identification of highly-forbidden optical transitions in highly charged ions}

\author{Shuying Chen$^{1,*}$, Lukas J. Spieß$^{1,*}$, Alexander Wilzewski$^1$, Malte Wehrheim$^1$, Kai Dietze$^1$, Ivan Vybornyi$^2$, Klemens Hammerer$^2$, José R. Crespo López-Urrutia$^3$, and Piet O. Schmidt$^{1,4}$}

\affiliation{$^1$QUEST Institute for Experimental Quantum Metrology, Physikalisch-Technische Bundesanstalt, Bundesallee 100, 38116 Braunschweig, Germany }
\affiliation{$^2$Institute of Theoretical Physics, Leibniz Universität Hannover,  Appelstrasse 2, 30167 Hannover, Germany}
\affiliation{$^3$Max-Planck-Institut für Kernphysik, Saupfercheckweg 1, 69117 Heidelberg, Germany }
\affiliation{$^4$Institut für Quantenoptik, Leibniz Universität Hannover, Welfengarten 1, 30167 Hannover, Germany}
\email{shuying.chen@quantummetrology.de\\
lukas.spiess@quantummetrology.de}
\vspace{10pt}

\maketitle


\section{ODF theory}
For simulation of the ODF signal for \HCI{Ca}{14}, two electronic states ($| g \rangle$ and $| e \rangle$) and 20 Fock states ($| n \rangle$) for the motional states are included via tensor calculation. The ion is assumed to be initially in the electronic ground state  $| g \rangle$ as well as the motional thermal state with $ \Bar{n}=0.05$.
The effective interaction Hamiltonian is $H = H_{\mathrm{car}} + H_{\mathrm{sb}}$, including the carrier part $H_{\mathrm{car}}$ which only couples the electronic states and the sideband coupling part $H_{\mathrm{sb}}$ coupling the electronic and first-order motional states. The two parts are given as \cite{leibfried_quantum_2003},
\begin{equation}
\begin{split}
    H_{\mathrm{car}} &= \dfrac{1}{2} \Omega_1 \sigma_{+} e^{i \delta t} + h.c., \\
    H_{\mathrm{sb}} &=  \dfrac{1}{2} \Omega_s \sigma_{+} \hat{a}^{\dag} e^{i (\delta \pm \omega_m) t} + h.c.,
\end{split}
\end{equation}
where $\hat{a}^{\dag}$ is the creation operator for motional states and its complex conjugate $ \hat{a}$ is the annihilation operator. $\sigma_+ = |e\rangle \langle g| $ is the coupling of the electronic states. $\Omega_1$ is the Rabi frequency for the electronic state coupling and $\Omega_s$ is the sideband Rabi frequency, which is given by \cite{leibfried_quantum_2003},
\begin{equation}
    \Omega_s = \eta \Omega_2 \exp(-\eta^2/2)\sqrt{1/(n+1)} L(n, 1, \eta^2). 
\end{equation}
$\Omega_2$ is the Rabi frequency of beam 2 coupling the carrier transition of the electronic states. Both $\Omega_1$ and $\Omega_2$ in the simulation are set to the experimentally measured values from Rabi flopping on the carrier transition to be around $2\pi\times 5$ kHz.  $L(n, 1, \eta^2) $ is generalized Laguerre polynomial \cite{leibfried_quantum_2003}. 

The dynamic evolution of the system is described by the Lindblad equation \cite{Intravaia_quantum_2003,Simon_measuring_2020},
\begin{equation}
    \dot{\rho} = - \frac{i}{\hbar}[H,\rho]+ \frac{1}{2} \sum_k (2 \hat{L}_k \rho \hat{L}_k^{\dag} - \hat{L}_k^{\dag} \hat{L}_k\rho - \rho\hat{L}_k^{\dag}\hat{L}_k),
    \label{eq:odf1}
\end{equation}
where $\rho$ is the density matrix of the states with 2 $\times$ 20 dimensions. $\hat{L}_k$ is the Lindblad operator, which includes the electronic state spontaneous decay, 
$\hat{L}_{\mathrm{spon}}$, and the motional state coupling to the environment, $\hat{L}_{\mathrm{env}}$. The coupling of the motional states to the environment causes heating and dephasing of the system,  $\hat{L}_{\mathrm{env}}=\hat{L}_{\mathrm{heating}}+\hat{L}_{\mathrm{dephasing}}$. No influence from the relative dephasing between the two beams has been observed experimentally, thus it is not included in the model. The spontaneous decay is given by
\begin{equation}
    \hat{L}_{\text{spon}} = \sqrt{\Gamma} | g \rangle  \langle e |.
\end{equation}
Here, $\Gamma = 1/\tau$ with the excited state lifetime $\tau$ of around 10~ms.
%
The motional heating process, considering a large number of reservoir phonons, can be described as \cite{leibfried_quantum_2003, Intravaia_quantum_2003, Simon_measuring_2020, turchette_decoherence_2000},
\begin{equation}
    \hat{L}_{\text{heating}} = \sqrt{\tilde{\gamma_1}} (\hat{a} + \hat{a}^{\dag}) .
\end{equation}
Here $\hat{a}$ corresponds to phonon damping, and $\hat{a}^{\dag}$ corresponds to the phonon heating.
$\tilde{\gamma_1}$ is the heating rate \cite{leibfried_quantum_2003, Intravaia_quantum_2003, Simon_measuring_2020}. 
The motional state dephasing is described as \cite{flühmann_encoding_2019},
\begin{equation}
    \hat{L}_{\text{dephasing}} = \sqrt{\tilde{\gamma_2}} (\hat{a}\hat{a}^{\dag} + \hat{a}^{\dag}\hat{a}) .
\end{equation}
Here $\tilde{\gamma_2}$ is the dephasing rate \cite{flühmann_encoding_2019}. 

Eq.~(\ref{eq:odf1}) is then numerically solved with the python package Qutip \cite{qutip} to get $P_{\text{ODF}}$. We find that $\tilde{\gamma_1} = \beta = 8.2$ and $\tilde{\gamma_2} = 150$~Hz reproduces our signal in Fig.~2(b) in the main text well. 
Similar calculations are also applied to the simulations for Fig.~3 in the main text, including all the three Zeeman sublevels.

The heating background together with the decoherence from coupling of the motional modes to the environment limits the efficiency or signal to noise ratio (SNR) of the ODF method when applying longer probe time. 



\section{LCS theory for Ca$^{+}$ transition scanning}
\begin{figure}[!htbp]
    \centering
    \includegraphics{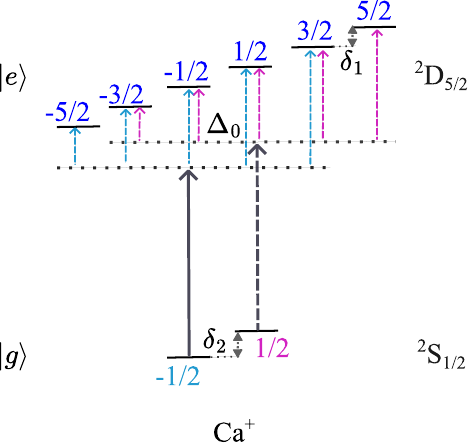}
    \caption{The coupled states of Ca$^{+}$ in the LCS interaction. }
    \label{fig:40Ca+}
\end{figure}
The interaction of the atomic states in Ca$^{+}$ with the scanning laser is expressed by the atomic density matrix $\hat{\rho}$. Its evolution is obtained using the Lindblad equation, Eq.~(\ref{eq:odf1}).

The interaction Hamiltonian is given by
\begin{equation}
H = \frac{1}{2} \sum_n \Omega_{n}e^{i\Delta_n t} (|e_{\mathrm{i}}\rangle \langle g_{\mathrm{k}}|)_\mathrm{n} + h.c. .
\label{hamit}
\end{equation}
Here $\Omega_{n}$ ($n=|\text{S},-1/2\rangle \rightarrow |\text{D},-5/2\rangle, |\text{D}, -1/2\rangle \rightarrow |\text{D},-3/2\rangle ...$) is the Rabi frequency of the corresponding coupled states from ground state $|g_{\mathrm{k}}\rangle$ ($|\text{S},-1/2\rangle$ and $|\text{S},1/2\rangle$) to excited state $|e_{\mathrm{i}}\rangle$ ($|\text{D},-5/2\rangle$...  $|\text{D},5/2\rangle$), as shown in Fig.~\ref{fig:40Ca+}. The Rabi frequencies used in Eq.~(7) are experimentally characterized with Rabi flops and are given in Table \ref{tab: LCS_Ca+}.
\begin{table}[]
    \centering
    \begin{tabular}{c|c|c|c}
    \hline
    \hline
         $m_J\rightarrow m_J'$ & $-1/2\rightarrow-5/2$ & $-1/2 \rightarrow -3/2$& $-1/2 \rightarrow -1/2$  \\
         
         $\Omega/2\pi/$ kHz& 8.76& 1.38& 13.91  \\
         \hline
     $m_J\rightarrow m_J'$ & $-1/2 \rightarrow 1/2$ & $-1/2 \rightarrow 3/2$ &   \\
         
         $\Omega/2\pi/$ kHz & 1.80 &3.05 &   \\
         \hline
         
         $m_J\rightarrow m_J'$ & $1/2 \rightarrow -3/2$& $1/2 \rightarrow -1/2$& $1/2 \rightarrow 1/2$  \\
         $\Omega/2\pi/$ kHz& 5.22 & 1.01& 11.40 \\
        \hline
      $m_J\rightarrow m_J'$ &  $1/2 \rightarrow 3/2$& $1/2 \rightarrow 5/2$ & \\
         $\Omega/2\pi/$ kHz& 3.52& 6.06 & \\
\hline
   \hline      
    \end{tabular}
    \caption{Rabi frequencies used in the \HCI{Ca}{} LCS experiment.}
    \label{tab: LCS_Ca+}
\end{table}
$\Delta_n$ is the detuning of the laser to the corresponding coupled states, and $\Delta_n = \Delta_0 + \varepsilon t +\delta_r$. $\Delta_0$ is the initial detuning of the laser frequency to transition $|\text{S},-1/2\rangle \rightarrow |\text{D},-1/2\rangle$, and $\delta_r$ is the frequency difference of other transitions relative to the transition $|\text{S},-1/2\rangle \rightarrow |\text{D},-1/2\rangle$.
$\varepsilon$ is the scan speed of the laser frequency.
$\delta_1 = 2 \pi \times 6~\rm{MHz}$ is the Zeeman splitting in the excited-state Zeeman sublevels, and $\delta_2 = 2 \pi \times 10~\rm{MHz}$ is the Zeeman splitting in the ground-state Zeeman sublevels. 

The Lindblad term $\hat{L_k} = \hat{L}_{\rm{decay}}(\hat{\rho}) + \hat{L}_{\rm{laser}}( \hat{\rho})$ is the total relaxation matrix. Here $L_{\rm{decay}}$ is a decay term given by
\begin{equation}
\centering
    L_{\rm{decay}} =  \sqrt{\Gamma} \sum_n|(g_{\mathrm{k}}\rangle \langle e_{\mathrm{i}}|)_n,
    \label{decay1}
\end{equation}
with the decay rate $\Gamma=1/\tau$ and $\tau$ the effective lifetime of the excited state. For the simulations in the main text, we used $\Gamma = 6.25$~Hz. This is faster than the natural spontaneous decay of the excited state due to incomplete extinction of repumper light in our system, which depopulates the excited state through the P$_{3/2}$ state to the ground state faster. The decay probability to the two ground-state sub-levels is assumed to be equal.
And $L_{\rm{laser}}$ is a dephasing term caused by the finite laser linewidth, with the Lindblad operator written as
\begin{equation}
    L_{\rm{laser}}= \sqrt{\gamma} \sum_n(|g_{\mathrm{k}}\rangle \langle g_{\mathrm{k}} |+|e_{\mathrm{i}}\rangle \langle e_{\mathrm{i}} |)_n, 
    \label{decay}
\end{equation}
where the sum runs over all the ground- and excited-state Zeeman components.

Finally the density matrix evolution of the states in Eq.~(\ref{eq:odf1}) can be solved numerically with the Python package Qutip \cite{qutip} to get the population transfer from the ground to the excited state. 
For the simulations in the main text, we set $\gamma/2\pi$ to 100~kHz. This laser linewidth term is essential to match the simulation results to the data.

\section{LCS theory for \HCI{Ni}{12} transition scanning}
\begin{figure*}[!htbp]
    \centering
    \includegraphics{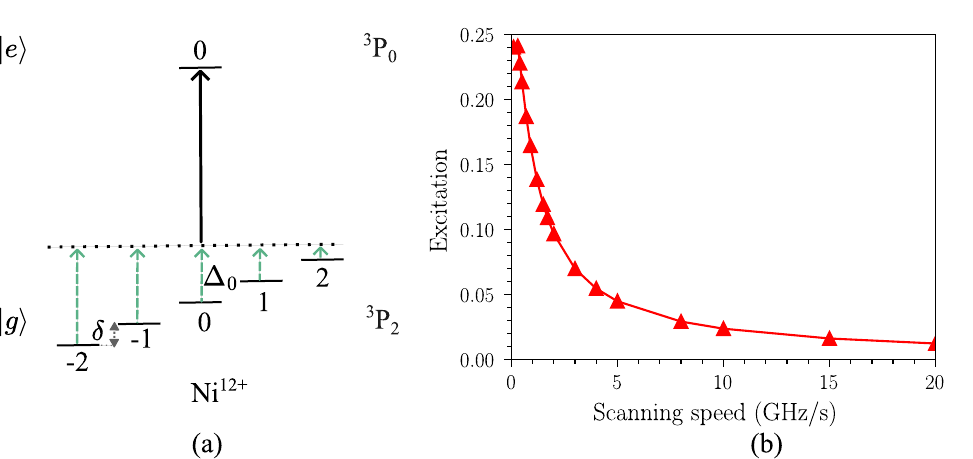}
    \caption{(a) The coupled states of \HCI{Ni}{12} in the LCS interaction. (b) The excitation probability for scanning across the \HCI{Ni}{12} clock resonance for different scan speeds using the LCS method. Triangles are the simulations and the solid line connecting the points is a guide to the eye. }
    \label{fig:Ni12+}
\end{figure*}
The interaction of the atomic states in \HCI{Ni}{12} with the scanning laser is expressed by the density matrix operator $\hat{\rho}$. Its evolution is obtained using the Lindblad equation in the form of Eq.~(\ref{eq:odf1}) using the Hamiltonian in the form of Eq.~(\ref{hamit}).
Here $\Omega_{n}$ ($n=|^3\text{P}_2, -2\rangle \rightarrow |^3\text{P}_0, 0\rangle, |^3\text{P}_2, -1\rangle \rightarrow |^3\text{P}_0, 0\rangle ...$) is the Rabi frequency of the corresponding coupled states. The corresponding levels are shown in Fig.~\ref{fig:Ni12+}(a). The initial state is assumed to be the ground state $|^3\text{P}_2, m_j = -2\rangle$, e.g. prepared through quantum logic-assisted optical pumping using the logic transition \cite{micke_coherent_2020}. We assume a Zeeman splitting of $\delta/2 \pi = 
500~\rm{kHz}$ in the ground-state sublevels, which can be tuned via the magnetic field. $\Delta_n=\Delta_0+\varepsilon t + \delta_{r}$ is the detuning of the applied laser frequency to the different transitions, and $\Delta_0$ is the initial detuning of the laser frequency relative to the resonant frequency of the transition $|^3\text{P}_2, 0\rangle \rightarrow |^3\text{P}_0, 0\rangle$. $\delta_{r}$ is the relative frequency difference between the other transitions and the transition $|^3\text{P}_2, 0\rangle \rightarrow |^3\text{P}_0, 0\rangle$.
$\varepsilon$ is the scan speed of the laser frequency and the scan direction is chosen to be such that the transition $|^3\text{P}_2, -2\rangle \rightarrow |^3\text{P}_0, 0\rangle$ is excited last to avoid depumping of the excited state.

The Lindblad equation in Eq.~(\ref{eq:odf1}) with similar terms as in Eq.~(\ref{decay1}) and Eq.~(\ref{decay}) above is solved numerically with the python package Qutip \cite{qutip} to get the excitation probability to the $|^3\text{P}_0\rangle$ state, and the result is shown in Fig.~\ref{fig:Ni12+}(b). The triangles are simulation results and solid line connects the points as a guide to the eye. This is then used to derive the effective speed in the main text in Fig.~5(c). 
The Rabi frequencies $\Omega_n$ are assumed to be maximally $5$~kHz , and the Rabi frequencies of the other transitions are calculated for optimised laser polarization such that the $|^3\text{P}_2, -2\rangle \rightarrow |^3\text{P}_0, 0\rangle$ transition strength is maximised
\cite{roos_thesis}. 
We assume $\Gamma = 1/\tau$ with the excited state lifetime $\tau$ of $20$~s, which is from the calculated natural lifetime \cite{yu_highly_2023}, and $\gamma/2\pi$ is $100$~kHz.

\end{document}